\RequirePackage{fix-cm}
\documentclass[smallextended]{svjour3}

\smartqed

\usepackage{graphicx}
\usepackage[margin=2.5cm, centering]{geometry}
\usepackage{mathptmx}
\usepackage{float}
\usepackage{caption}
\usepackage[english]{babel}
\usepackage{csquotes}
\usepackage[compress]{cite}
\usepackage{rotating}
\usepackage{afterpage}
\usepackage{hyperref}

\usepackage{amssymb}

\begin{document}

\title{A tutorial on THz pulse analysis: accurate retrieval of pulse arrival times, spectral energy density and absolute spectral phase.}

\author{James Lloyd-Hughes \and Nishtha Chopra \and Justas Deveikis \and Raj Pandya \and Jack Woolley
}

\institute{	J.\,Lloyd-Hughes, N.\,Chopra, J. Deveikis \at
	Department of Physics, University of Warwick, Coventry, United Kingdom 
    \email{j.lloyd-hughes@warwick.ac.uk}
	\and
    J. Woolley \at
	Warwick Centre for Ultrafast Spectroscopy, University of Warwick, Coventry, United Kingdom
	\and
	R.\ Pandya \at
	Department of Chemistry, University of Warwick, Coventry, United Kingdom
 }

\date{Received: date / Accepted: date}

\newcommand{\sech}{\mathrm{sech}}
\newcommand{\eg}{\emph{e.g.}\ }
\newcommand{\ie}{\emph{i.e.}\ }
\newcommand{\tCEP}{t_{\mathrm{CEP}}}

%\usepackage{natbib}
%\setcitestyle{super}

\maketitle

\begin{abstract}
Electro-optic sampling allows the electric field of THz, mid-infrared and visible light pulses to be measured directly as a function of time, with data analysis often performed in the frequency domain after fast Fourier transform. 
Here we review aspects of Fourier theory relevant to the frequency-domain analysis of light pulses recorded in the time-domain.
We describe a ``best practise'' approach to using the discrete Fourier transform that ensures consistency with analytical results from the continuous Fourier transform.
We summarise a phenomenological time-domain model of THz pulses, based on carrier and envelope waves, and show that it can reproduce a wide variety of experimental single- to multi-cycle THz pulses, with exemplary data from lab-based sources (photoconductive antennae, optical rectification, spintronic emitters) and a THz free electron laser.
A quantitative comparison of the spectral energy density of these distinct sources is enabled by the amplitude-accuate discrete Fourier transform.
We describe a method that ensures the accurate calculation of the absolute spectral phase (valid for arbitrary sampling windows in the time-domain) and summarise how the carrier-envelope phase, pulse arrival time and chirp can be obtained from the phase.
Our aim with this overview of THz pulse analysis is to highlight algorithms and concepts that are useful to newcomers to time-domain spectroscopy and experts, alike.
\end{abstract}

\section{Introduction}

In time-domain ultrafast spectroscopy the electric field of a pulse of electromagnetic radiation, $E(t)$, is sampled experimentally as a function of time, and then Fourier transformed to obtain the complex spectra $\widetilde{E}(\omega)=|\widetilde{E}(\omega)|e^{i\phi(\omega)}$  at angular frequency $\omega$. 
The amplitude spectrum, $|\widetilde{E}(\omega)|$, represents the frequency content, while the phase $\phi(\omega)=\angle \widetilde{E}(\omega)$ marks the arrival time of each frequency component relative to zero time, where $\angle$ denotes calculating the angle of the complex number. 
The well-established technique of terahertz time-domain spectroscopy (THz-TDS) makes full use of electro-optic sampling or photoconductive detection to record $E(t)$ after interacting with a sample, where knowledge of the amplitude and phase gives the complex refractive index (complex dielectric function) of the material of interest \cite{Lloyd-Hughes2012}.
While demonstrated first for THz pulses, time-domain spectroscopy has been widely utilised in the mid-infrared \cite{Huber2000} and recently even in the visible (or ``petahertz'') range \cite{Zimin2021,Herbst2022}.

Although often only the amplitude spectrum is reported, the spectral phase, $\phi(\omega)$, contains a wealth of information about an electromagnetic pulse, including the pulse's arrival time (from the group delay) and the carrier-envelope phase of the pulse, which creates a frequency-independent offset to the spectral phase \cite{Kawada2016}.
Further, the phase provides information about dispersive processes that occur during pulse generation, propagation, or detection.
Propagation through a conductive or absorptive material can alter the spectral phase of a pulse, via its frequency-dependent refractive index \cite{Lloyd-Hughes2012}.
Changes can even occur during propagation in free space: broadband electromagnetic pulses change spectral phase as a result of the geometric aberrations of the optical system, which alter the path length \cite{Chopra2023}; the Gouy effect causes the carrier-envelope phase to evolve as a beam propagates through a focus \cite{Kuzel1999}.
Schemes that modify the carrier-envelope phase of a THz pulse have been reported \cite{Kawada2016,Allerbeck2023}, and are particularly important when using intense THz electric fields to drive non-linear processes, such as in THz scanning tunneling microscopy \cite{Lloyd-Hughes2021, Allerbeck2023}.

In the vast majority of spectroscopy studies, rather than analysing complex spectra $\widetilde{E}(\omega)$, analysis is instead performed on ratios of two different spectra, $\widetilde{E}_1(\omega)$ and $\widetilde{E}_2(\omega)$. 
Such ratios are assumed to cancel (to a first approximation) any changes in spectral phase as a result of propagation in the spectrometer or the detection process.
For example, in transmission spectroscopy the complex transmission $\widetilde{T}(\omega)=\widetilde{E}_2/\widetilde{E}_1$ for sample spectrum $\widetilde{E}_2$ and reference spectrum $\widetilde{E}_1$ is analysed, including calculating the phase difference between $\widetilde{E}_1$ and $\widetilde{E}_2$, $\Delta\phi(\omega)=\phi_2-\phi_1$.
This \emph{relative} phase provides information about differences in the optical path length as a result of the introduction of a sample into the THz beam  \cite{Lloyd-Hughes2012}.
However, the \emph{absolute} phases $\phi_1(\omega)$ and $\phi_2(\omega)$ are rarely reported. 

Challenges associated with the accurate retrieval of the relative phase $\Delta\phi(\omega)$ in THz-TDS have been discussed previously. 
In the context of transmission THz-TDS, Jepsen discussed difficulties in unwrapping the relative phase of sample and reference spectra, and provided an algorithm that made relative phase retrieval more robust by subtracting a linear phase term from sample and reference spectra  \cite{Jepsen2019}.
In the reflection geometry, the extra path length that arises when samples are not placed at exactly the same location along the beam direction leads to a phase ``error'', which can be corrected either experimentally by accurately positioning the reference mirror, or computationally \cite{Vartiainen2004}.
While these works identified how to improve analysis of the relative phase, they did not provide a way to reliably recover and utilise the absolute phase of the electromagnetic pulse. 

In this tutorial we explore the absolute phase of electromagnetic pulses, using exemplary time-domain waveforms and complex spectra from THz-TDS. 
We show that care must be taken to accurately calculate the absolute phase spectrum when using the discrete Fourier transform from popular numerical libraries.
Herein, absolute phase refers to the phase calculated with respect to the centre of the electromagnetic pulse, and which is independent of the experimental parameters (e.g.\ the start of the sampling window, or the number of points).
We provide a definition of the discrete Fourier transform (DFT) that correctly determines the amplitude and absolute phase spectra, and verify it by comparsion with analytical results from continuous Fourier transform (CFT) theory.
We then construct a simple mathematical picture of free-space electromagnetic pulses in the time-domain that allows single- or multi-cycle waveforms to be fit to experimental data, either in the time-domain or the frequency-domain, for arbitrary carrier-envelope phases and pulse arrival times.
Our accurate Fourier theory approach further allows us to:
\begin{itemize}
 \item determine a unique arrival time for a pulse. For a few- or single-cycle pulse, as common in THz-TDS, this can be at the central zero of the electric field, rather than the peak electric field;
 \item compare the spectral energy density of disparate pulsed THz sources (a free electron laser, and laser-based sources using a spintronic emitter, optical rectification or a photoconductive antenna);
 \item examine the absolute phase of experimental waveforms: for example, finding the carrier-envelope phase, or the chirp created during electro-optic sampling;
 \item identify that the exponential dependence of amplitude spectra in broadband THz pulses is linked to sech-like pulse envelopes.
\end{itemize}

The article is structured as follows. 
Section II provides a robust and accurate DFT routine that produces the same amplitude and phase spectra for simple waveforms as found analytically from the continuous Fourier transform.
Care is taken to try to provide a ``best practice'' approach to using the DFT that avoids pitfalls in recovering the absolute phase when applied to typical waveforms, which have asymmetric sampling windows in the time-domain. 
Section III then describes the application of Fourier theory to construct a model of electromagnetic pulses, including a discussion of the carrier-envelope phase and the pulse arrival time (group delay), valid for narrowband (multi-cycle) or broadband (single-cycle) pulses. 
The analysis of experimental data from THz-TDS in Section IV illustrates how these Fourier theory results can be used in practise.
To avoid impeding the flow of the main text appendices provide useful results from Fourier theory, a python implementation of the DFT, and a description of how to deconvolve the electro-optic response function. 
The numerical routines we developed here are available online \cite{dft-github}.

\section{Ensuring amplitude- and phase-accuracy with the DFT}
\label{sec:dft}
To begin our tutorial on the use of Fourier theory to analyse time-domain data of light pulses, we first discuss and define the expressions used to obtain the CFT and the DFT. The CFT can be used to calculate the spectrum of a real, continuous function $E(t)$, and is defined as $\widetilde{E}(\omega) = \int_{-\infty}^{\infty} E(t) e^{i\omega t} \,dt$, following the sign convention adopted in physics and optics.
In experiments, however, $E(t)$ has to be sampled discretely, yielding a sequence $E_m(t_m)$, where $m=0, 1, 2, ... n-1$ for $n$ sampling points, and
the $m$th sample is at time $t_m =m \delta t$.
The width of the whole time window is $T=n \delta t$.
As $E_m$ is a discrete series, the discrete Fourier transform (DFT) has to be used to obtain the spectrum, defined by
$\widetilde{E}_k(\omega_k) = \sum_{m=0}^{n-1} E_m(t_m) e^{i \omega_k t_m}$, where the complex spectral component $\widetilde{E}_k$ is sampled at each angular frequency $\omega_k=2\pi k/(n\delta t)$ for $k=0,1,2...n-1$.
%In THz-TDS, $E(t)$ corresponds to the electric field of a THz pulse, and the experimental samples $E_m$ at different times are often obtained by moving a motorised stage to change the optical path length by a time $\delta t$ between discrete sampling points.
Note that if the experimental samples are not evenly spaced in time, then the data must first be interpolated onto a time axis with uniform spacing.
%Further, by taking the discrete Fourier transform we assume that the signal is periodic with period $T$. 
%Hence the forward DFT defined above corresponds to the inverse DFT of these packages.

For functions $E(t)$ where the CFT can be found analytically, the DFT calculated by computer should agree with the analytical CFT result. 
We now use this reasoning to provide a phase- and amplitude-accurate DFT.

\subsection{Derivation}
By comparing the definitions of the CFT and the DFT we can see that in order for their amplitude spectra to match we require $|\widetilde{E}_k| \simeq |\widetilde{E}(\omega=\omega_k)|$, and hence the DFT definition, as stated above, is missing the term $\delta t$. 
We therefore define the DFT as
%\tcbhighmath[drop fuzzy shadow]
\begin{equation}
{\widetilde{E}_k = \delta t \sum_{m=0}^{n-1} E_m e^{i \omega_k t_m}}
 \label{eq:DFT3}
\end{equation}
in order to be consistent with the CFT. 
Parseval's theorem (Appendix \ref{app:Fourier}, Eqn.\ \ref{eq:Parseval}) can also be employed to check that the frequency-domain amplitude is correctly scaled, by ensuring that the time-domain and frequency-domain integrals match. While ensuring correct amplitude scaling for the DFT is well known, we highlight its importance for the THz community in Section \ref{sec:exp} by comparing the amplitude spectra and spectral energy density of disparate THz sources: photoconductive antennas driven by laser oscillators, a spintronic emitter or optical rectification in LiNbO$_3$ excited by pulses from an amplified laser, and multi-cycle THz pulses produced by a THz free electron laser.

\begin{figure}[tb]
	\centering
	\includegraphics[width=1.0\columnwidth]{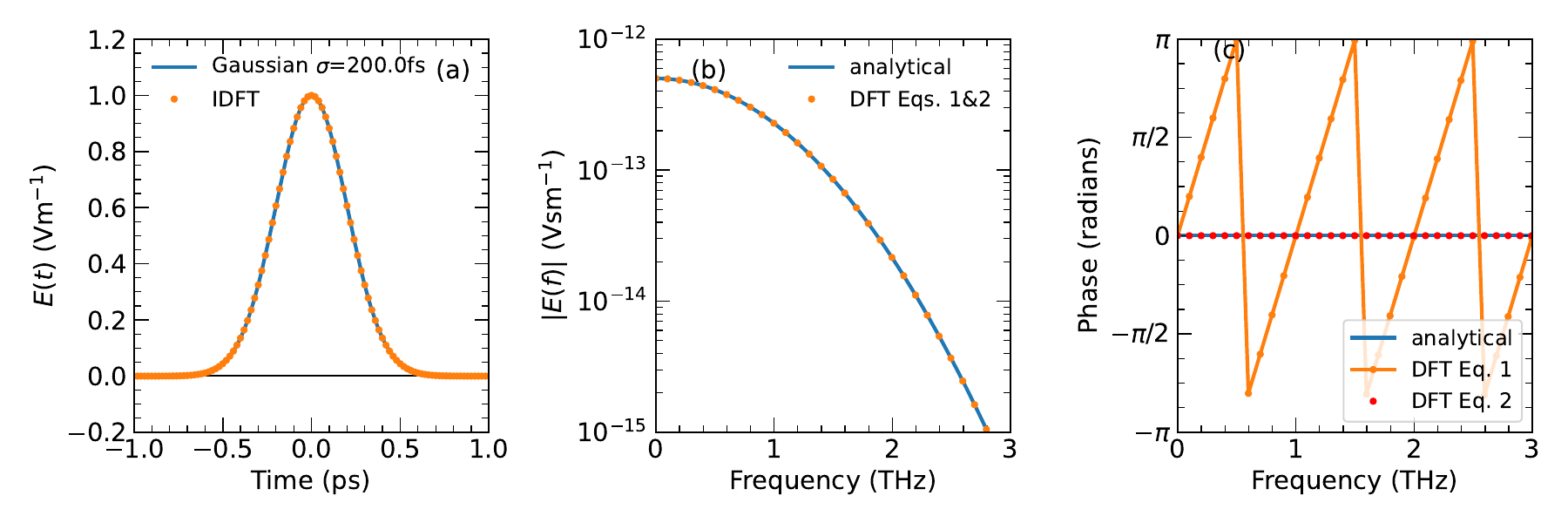}
	\caption{(a) A Gaussian time-domain pulse (solid line), with $\sigma=200$\,fs, sampled with 501 points from $t=-1$\,ps to 9\,ps, and centred at zero time, at $T_0=1$\,ps after the first point in the discrete time series. Points show the inverse DFT calculated from the spectra using Eqn.\ \ref{eq:IDFT}. (b) The amplitude spectra of (a) calculated from the CFT (solid line) and DFT from Eq.\ \ref{eq:DFT3} and Eq. \ref{eq:DFT5} (points). (c) The phase spectra from the CFT (blue line, at zero phase), the uncorrected DFT (Eq.\ \ref{eq:DFT3}, orange) and phase-corrected DFT (Eq. \ref{eq:DFT5}, red points). The uncorrected DFT has a linear slope with gradient $2\pi T_0$, whereas the phase-corrected DFT reproduces the zero phase expected analytically.}
	\label{fig:correctDFT}       
\end{figure} 

Although the mathematical form of the DFT and CFT now look similar, there is an additional subtlety hidden in requiring that the phase returned by the DFT should match that of the CFT, or $\angle \widetilde{E}_k=\angle \widetilde{E}(\omega=\omega_k)$. 
This requirement is not ensured by Equation \ref{eq:DFT3}, which is ``blind'' to the absolute time $t$. 
The DFT as stated above (and in standard numerical libraries \cite{ScipyMatlab}) considers the time $t_m=m\delta t$ relative to the first time point sampled in the series, at $m=0$.
Hence the phase $\phi_k=\angle \widetilde{E}_k$ using Equation 1 is determined relative to that of the first time point in the series, rather than reporting the absolute phase $\phi(\omega=\omega_k)$ relative to the pulse's centre.
In experiments measuring a pulse this definition is therefore unsatisfactory, as the relative phase $\phi_k$ returned from the standard DFT varies with the length of the sampling window before the pulse.
To determine the absolute phase relative to the pulse centre we define the DFT as:
\begin{equation}
%\tcbhighmath[drop fuzzy shadow]
{\widetilde{E}(\omega_k) = \delta t \sum_{m=0}^{n-1} E_m e^{i \omega_k (t_m-T_0)} = \delta t \cdot e^{-i\omega_k T_0} \sum_{m=0}^{n-1} E_m e^{i \omega_k t_m} }
 \label{eq:DFT5}
\end{equation}

\noindent where the term $e^{-i \omega_k T_0}$ accounts for the finite arrival time of the pulse at element $m_0$ and time $T_0=m_0\delta t$ after the first time point, and the factor $\delta t$ gives the correct spectral amplitude. 
Note that throughout we adopt $\widetilde{E}(\omega_k)$ to refer to the phase- and amplitude-correct DFT of Equation \ref{eq:DFT5}, and use $\widetilde{E}_k$ to denote the DFT of Equation \ref{eq:DFT3}, which has the correct amplitude but an inaccurate phase.

A derivation of the phase-corrected DFT equation is provided in Appendix \ref{app:DFT}, while an example of its numerical implementation is provided in the Python code presented in Appendix \ref{app:code}. 
Here we establish its validity by comparing the discrete spectrum with the result from the analytical CFT for the case of a Gaussian input pulse, centred at zero time and with the form $g(t)=A e^{-t^2/2\sigma^2}$, as pictured in Fig.\ \ref{fig:correctDFT}(a) for $A=1$ and $\sigma=200$\,fs. 
The amplitude and phase of $\widetilde{g}(\omega)$ from the CFT are shown in Fig.\ \ref{fig:correctDFT}(b) and (c) [solid blue lines]. 
Since $g(t)$ is a real, even function, the CFT $\widetilde{g}(\omega)=\int_{-\infty}^{\infty}g(t) e^{i\omega t}dt = \sqrt{2 \pi \sigma^2} e^{-\omega^2 \sigma^2 /2}$ is purely real, and therefore has zero argument, \ie zero phase.
While the amplitude-corrected DFT (Equation \ref{eq:DFT3}) applied to the same, discretely sampled Gaussian function $g_m$ yields the correct amplitude spectrum (points in panel (b)) it returns the wrong phase (orange points in panel (c)). 
Rather than zero, the phase is a straight line, and phase jumps of $-2\pi$ occur every time the phase increases above $\pi$ (these can be straightforwardly unwrapped if needed). 
In contrast, Equation \ref{eq:DFT5} returns the same amplitude (orange points in (b)) and also phase (red points in (c)) as the CFT, demonstrating that the definition of the DFT in Equation \ref{eq:DFT5} accurately reproduces the correct absolute phase in this case.

Note that the (inaccurate) linear phase returned from the standard DFT expression (Equation \ref{eq:DFT3}) should not be mis-interpreted as arising from anything physical: for example in THz-TDS one's physical intuition might be tempted to link this linear phase to light propagation by a distance $L$, since the phase of the electric field in that case is $\phi(\omega)=\omega L/c$.
Rather, the linear phase returned by Equation \ref{eq:DFT3} is a result of the pulse peaking at time $T_0=m_0\delta t$ from the first element ($T_0=1$\,ps in this case).
Since changing the sampling window to start earlier or later in time alters the phase calculated from Equation 1, it does not reproduce the expected zero phase of the CFT, nor does the linear phase represent anything physical.
There should be no difference in $\widetilde{E}(\omega)$ for data sampled with different time windows, however the phase using Equation 1 changes if $m_0$ or $\delta t$ changes.
Further, having a large linear phase can lead to significant phase unwrapping problems when the gradient is particularly high, such that phase jumps of $2\pi$ can occur within one step in frequency, as described in detail in Ref.\ \cite{Jepsen2019}.

\subsection{Inverse transform}
Similar considerations allow us to recommend a corrected inverse DFT that can convert a complex, frequency-domain signal into the time-domain. The inverse CFT is:
\begin{equation}
E(t) = \frac{1}{2\pi} \int_{-\infty}^{\infty} \widetilde{E}(\omega) e^{-i\omega t} \,d\omega,
\label{eq:icft}
\end{equation}

\noindent and the amplitude- and phase-corrected inverse DFT has the form 
\begin{equation}
%\tcbhighmath[drop fuzzy shadow]
{E_m(t_m) = \frac{1}{n \delta t} \sum_{k=0}^{n-1} \widetilde{E}_k e^{-i\omega_k t_m}}
\label{eq:IDFT}
\end{equation}

\noindent where $\widetilde{E}_k= \widetilde{E}(\omega_k) e^{i\omega_k T_0}$. The scaling factor $1/(n\delta t)$ ensures the correct amplitude, and can be derived by considering that $\delta \omega=\omega_{k+1}-\omega_k=2\pi / (n\delta t)$.
The validity of the phase-corrected inverse DFT is demonstrated in Fig.\ \ref{fig:correctDFT}(a), where the points show $E_m$ obtained from Eqn.\ \ref{eq:IDFT} applied to the complex spectrum of the Gaussian.

\subsection{Robustness of phase retrieval}

Because the DFT assumes that zero time corresponds to the first data point ($m=0$), it is common practise to re-order the input array to ensure an accurate representation of the phase, as an alternative to Equation 2.
This is typically done via a routine such as \texttt{ifftshift} in Python/NumPy or MATLAB, denoted here as ${S}^{-1}(E_m)$, which moves the right-hand half of the array $E_m$ to before the left-hand side. 
For example, ${S}^{-1}([-2,-1,0,1,2]) = [0,1,2,-2,-1]$.
After this shift, the first element of the array corresponds to the temporal centre of the pulse, aligning it correctly for the DFT if the pulse is centred at the mid-point of the array. 
\begin{figure}[t]
	\centering
	\includegraphics[width=0.9\columnwidth]{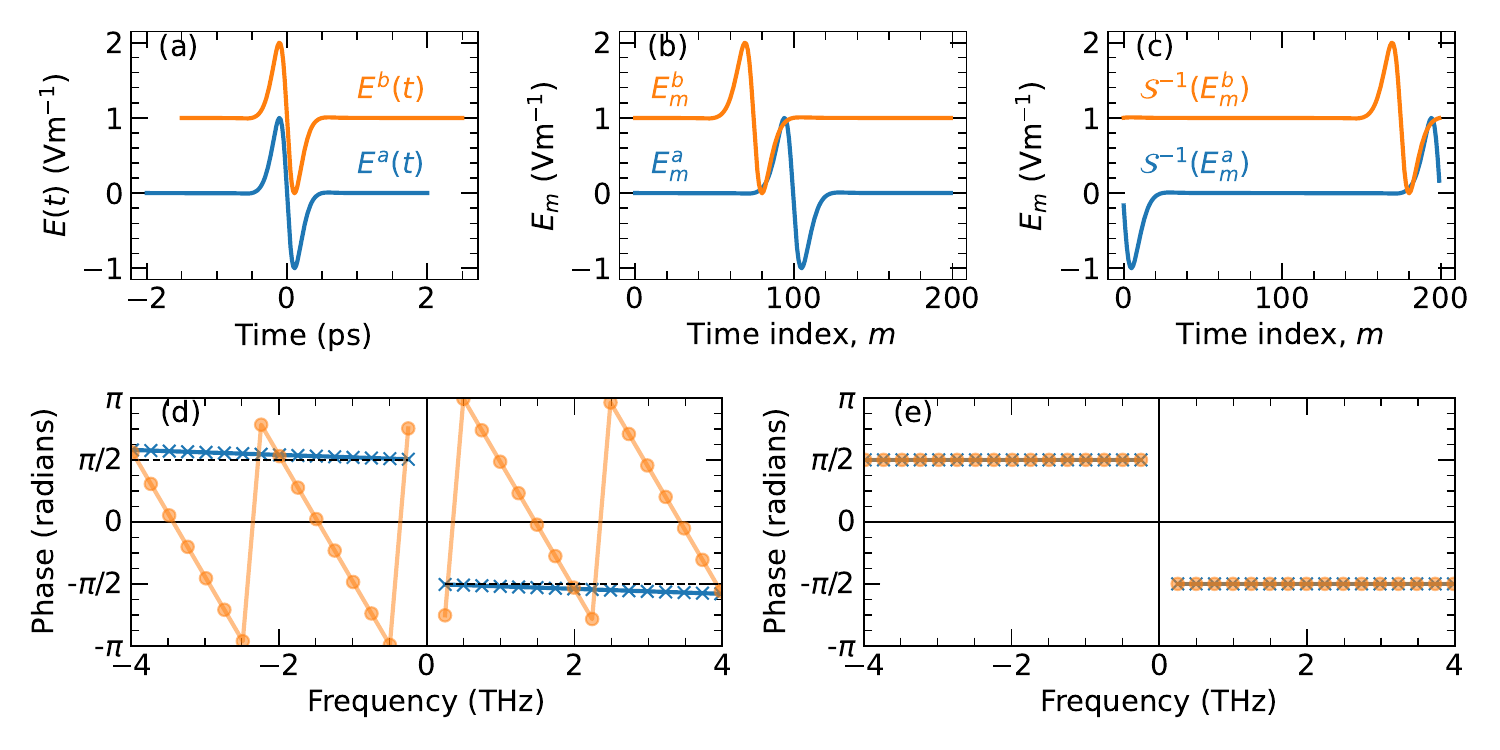}
	\caption{(a) Single-cycle THz pulse (Equation \ref{eq:THz-pulse-time-domain} with $A=1$, $t_e=\tCEP=0$, $\tau$=100\,fs) sampled from -2ps to 2ps with $N=200$ points (blue), and the same pulse sampled from -1.5ps to 2.5ps (orange, offset vertically for clarity). (b) Time series $E_m^{a,b}$ for the data in (a), as a function of index $m$. (c) Time series after applying the ifftshift algorithm. (d) Phase of DFT of time-shifted $E_m^a$ for $N=200$ (blue crosses), $N=201$ (black dashed line) and $E_m^b$ (orange points, $N=200$). 
	The symmetric time-window deviates from the expected $\pm \pi/2$ by a small linear term. The asymmetric time-window has a larger linear slope as the ifftshift algorithm did not put the pulse centre at index $m=0$. (e) Phase retrieved using Equation \ref{eq:DFT5} acting on the unshifted data in panel (b), illustrating accurate phase retrieval. }
	\label{fig:phase-accuracy}
\end{figure}

The typical procedure is illustrated in Fig.\ \ref{fig:phase-accuracy} for a single-cycle THz pulse (Eqn.\ \ref{eq:THz-pulse-time-domain}, $\tau=100$\,fs) sampled from -2ps to 2ps with 200 points (blue curve), and the same pulse sampled from -1.5ps to 2.5ps (orange, offset vertically for clarity). 
If the numerical procedure is robust and accurate, the expectation is that the phase retrieved should be the same, as the pulses are identical.
In Fig.\ \ref{fig:phase-accuracy}(b) both time series are shown as a function of index $m$, while in panel (c) the arrays have been reordered by \texttt{ifftshift}. 
The DFT of the shifted data set has a phase shown in Fig.\ \ref{fig:phase-accuracy}(d): while the phase is close to $\pm \pi/2$ for the case $E^a(t)$ with a symmetric time window (-2 to 2\,ps), it clearly differs from the ideal values (dashed black lines, from $N=201$) as a result of the small asymmetry in the numerical sampling window arising from the choice of an even number of elements. 
For the asymmetric time-window $E^b(t)$ the phase has a large linear gradient, as the pulse centre is not mapped to the first element by \texttt{ifftshift}. 
Hence, both an odd number of points and a symmetric time-window are required for the ``shift'' approach to achieve accurate phase retrieval.

In real experiments, however, a waveform is rarely at the centre of the data series. 
This could be because the experimentalist does not know exactly where the centre of the pulse is (for example if it has arbitrary carrier envelope phase, as described in Section \ref{sec:THz-CEP}), or because a “causal” time window has been adopted, where meaningful data about a sample is only obtained for times during and after the THz pulse. 
Fortunately, the phase retrieval from Equation \ref{eq:DFT5} is more reliable: an accurate phase is returned for even or odd $N$, and for symmetric or asymmetric time-windows, as illustrated in Fig.\ \ref{fig:phase-accuracy}(e).
The key requirement is knowledge of the pulse's arrival time, $T_0$, which we provide methods for determining in the next Section, along with a Fourier treatment of the properties of THz pulses.

\section{A simple model of a THz pulse}
\label{sec:THz}

In this Section we establish simple expressions for the time- and frequency-domain waveforms of single- and few-cycle electromagnetic pulses. 
This provides a straightforward parameterisation of a THz pulse that can model single- or multi-cycle THz pulses.
The electromagnetic pulses from common laser-based THz sources, such as photoconductive switches \cite{Mosley2017,Ou2024}, optical rectification \cite{Hoffmann2007,Hirori2011}, and spintronic emitters \cite{Seifert2016} are often close to single-cycle in shape.
Multi-cycle THz pulses can be obtained from optical rectification \cite{Mosley2023}, photoconductive antenna arrays \cite{Froberg1992,Weling1994}, quantum cascade lasers \cite{Barbieri2011}, or free electron lasers \cite{Green2016,Chen2022}.
The frequency-domain analytical results were verified by comparison to the phase-accurate DFT of the time-domain model.
By first discussing the analytical model in this section, with insights from our amplitude- and phase-correct Fourier theory, we provide a pedagological treatment of the expected changes in amplitude and phase spectra for typical pulsed THz sources on changing the properties of a THz pulse (e.g.\ carrier-envelope phase, group delay, group envelope), before providing a comparison with experiment in Section \ref{sec:exp}.

\subsection{Carrier-envelope model}
To provide a phenomenological time-domain model of a few-cycle THz pulse we consider the electric field to be the product of a carrier wave, described by the function $f_c(t)$, and an envelope function $f_e(t)$, \ie $E(t)=f_c(t) f_e(t)$.
Note that the following is not an attempt to model the pulse shape of THz radiation based on the physics of the generation mechanism in the source, nor does it consider the impact of the propagation of THz radiation through a material or an optical system. 
Rather, the goal in this section is to derive a simple, phenomenological picture of THz pulses in the time-domain and, via the corrected discrete Fourier transform or analytical CFT, in the frequency domain.
 
For the carrier wave we assumed $f_c(t)=-\sin (\omega_c (t-t_e) - \phi_0)$, where the envelope function and carrier wave are both centred at time $t_e$ when the carrier-envelope phase, $\phi_0=\omega_c(t_c-t_e)$, is zero. 
We consider two possibilities for the envelope function: a Gaussian envelope, $f_e(t)=A e^{-(t-t_e)^2/2\sigma^2}$, or a $\sech$ envelope, $f_e(t)=A \sech ((t-t_e)/\tau)$.
This yields, for the sech case: 
\begin{equation}
%\tcbhighmath[drop fuzzy shadow]
{E(t) = -A \sech ((t-t_e)/\tau) \sin(\omega_c (t-t_e) - \phi_0).\label{eq:THz-pulse-time-domain}}
\end{equation}

As discussed in Ref.\ \cite{Ahmed2014}, the carrier-envelope model does not produce realistic electric field pulses for short pulse envelopes ($\tau \leq 2\pi/\omega_c$) if the carrier-envelope phase is such that the carrier wave becomes close to a cosine (e.g.\ when $\phi_0=\pi/2$).
In this case the spectral amplitude is non-zero at zero frequency, which is not physical for electromagnetic pulses in the far-field (where the electric field depends on a time derivative, hence has no dc component) \cite{Ahmed2014}.
Here, the choice of a sine carrier wave allows pulses that are close to single-cycle to be modelled when the carrier-envelope phase offset $\phi_0$ is close to zero or $\pi$, as the spectral amplitude drops to zero at zero frequency.
This approach is valid when $t_c=t_e$, i.e.\ zero carrier-envelope phase, which we consider in this section, while a version of the model valid for all CEPs is presented in Sec.\ \ref{sec:THz-CEP}.

\begin{figure}[t]
	\centering
	\includegraphics[width=1.0\columnwidth]{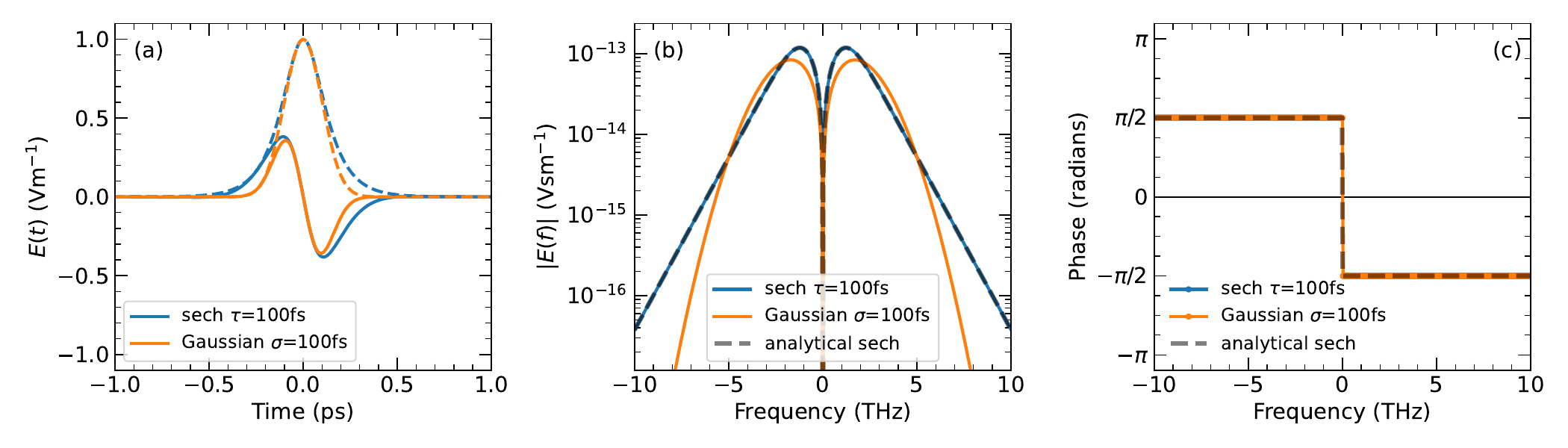}
	\caption{THz pulses calculated for sech (blue lines) or gaussian (orange lines) pulse envelopes. (a) Indicates the time-domain waveforms (solid lines) for a sech envelope (blue dashed line) or gaussian envelope (orange dashed line). A 1\,THz carrier frequency was assumed, with zero carrier and envelope offset times, and $\tau=\sigma=100$\,fs. (b) Amplitude spectra from the DFT of the THz pulses in (a), along with the analytical result for the CFT of a sech pulse (dashed line). (c) Phase spectra of data in (a) from DFT and CFT.}
	\label{fig:sech-v-gaussian}       
\end{figure} 

In Figure \ref{fig:sech-v-gaussian}(a) THz pulses $E(t)$ are shown with $A=1$Vm$^{-1}$ for the two envelope functions, with $\tau=\sigma=100$\,fs and $t_e=t_c=\tCEP=0$.
The THz pulse with the $\sech$ envelope (solid blue line) has broader wings in time than the THz pulse with the Gaussian envelope (solid orange line).
This is a result of the exponential time-dependence at large times of the sech pulse envelope (blue, dashed), which drops slower than a Gaussian envelope (orange, dashed).
In the frequency domain, the DFT of the THz pulses with sech and gaussian envelopes are reported in Fig.\ \ref{fig:sech-v-gaussian}(b) and (c), as calculated using equation \ref{eq:DFT5}.
The DFT of the sech pulse exhibits an amplitude spectrum (panel b, blue line) that decreases exponentially at high frequencies.
In contrast, $|E(\omega)|$ for the Gaussian envelope has a Gaussian spectral shape, and hence decreases more rapidly at high frequencies than the sech pulse.
Thus, despite the apparently longer time-domain envelope for the sech envelope, at higher frequencies (above 5\,THz for this particular case) sech pulses have larger spectral amplitude than Gaussian pulses.

The CFT of the model THz pulse $E(t)=f_c(t) f_e(t)$, valid for zero carrier-envelope phase ($\phi_0=0$), can be derived analytically using the standard Fourier theory expressions provided in Appendix \ref{app:Fourier}. 
For example, for a sech pulse envelope with $t_e=t_c=0$, using Equations \ref{eq:FTsech} and \ref{eq:product with sine} yields: 
\begin{equation}
 \widetilde{E}(\omega) = i A \frac{\pi \tau}{2} 
 \left[ \sech \left( \frac{\pi \tau}{2} \left( \omega+\omega_c \right) \right) -  \sech \left( \frac{\pi \tau}{2} \left( \omega-\omega_c \right) \right) \right]\label{eq:CFT-THz-pulse}
\end{equation}

\noindent where the THz pulse has a purely imaginary spectrum as a result of the time-domain being an odd function. 
The CFT result, shown by the dashed lines in Fig.\ \ref{fig:sech-v-gaussian}(b) and (c), is in excellent agreement with the DFT calculation from Equation \ref{eq:DFT5}.
The phase (Fig.\ \ref{fig:sech-v-gaussian}c) is $\pm \pi/2$, as the spectrum in Eqn.\ \ref{eq:CFT-THz-pulse} is purely imaginary ($e^{\pm i \pi/2}=\pm i$).
Further, the phase undergoes a $\pi$ change across zero frequency as a result of the two $\delta$-functions in Equation \ref{eq:FTsine} having opposite sign.

%It is worth noting that both positive and negative frequencies are plotted in Fig.\ \ref{fig:sech-v-gaussian}(b) and (c) in order to show the behaviour across $\omega=0$. 
%A mathematical justification for the need to include negative frequency components is that the CFT of a sine wave contains delta functions centred at both negative and positive frequencies.

\subsection{Influence of pulse duration: spectral slope}
\begin{figure}[bt]
	\centering
	\includegraphics[width=1.0\columnwidth]{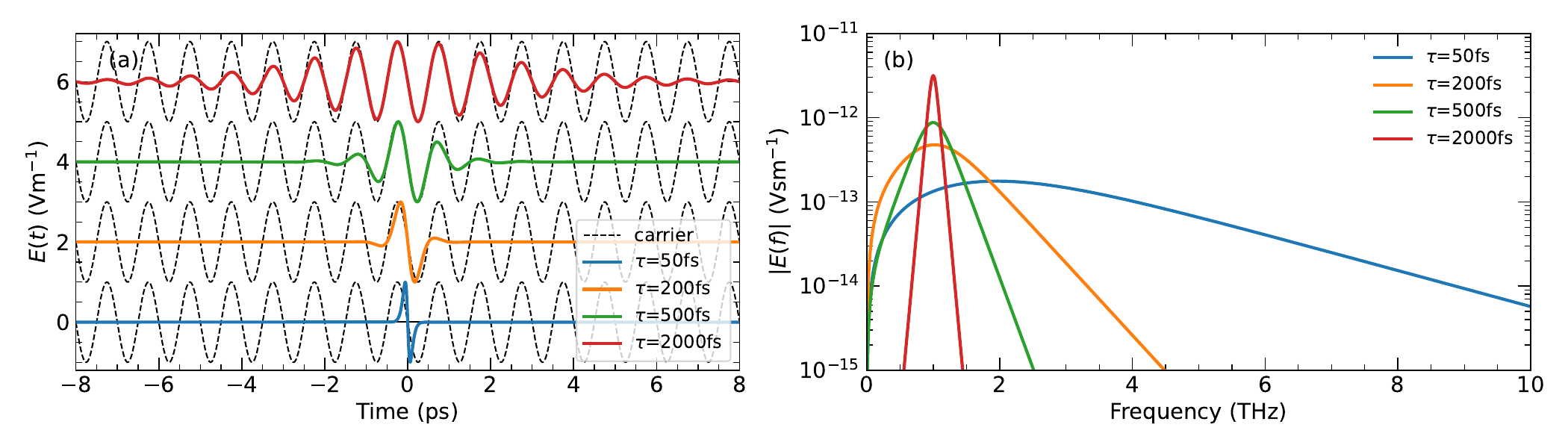}
	\caption{Impact of pulse duration. (a). Time-domain THz pulses for a $\sech$ envelope with different pulse durations, $\tau=50$\,fs to $2000$\,fs (coloured lines). Waveforms are offset for clarity. Dashed lines indicate the carrier wave (1\,THz frequency, or 1\,ps period). (b) Amplitude spectra of the THz pulses in (a).}
	\label{fig:sech-tau}       
\end{figure}

To demonstrate the impact of the envelope pulse's duration on the time- and frequency-domain, we report in Figure \ref{fig:sech-tau} results calculated using Equations \ref{eq:THz-pulse-time-domain} and \ref{eq:CFT-THz-pulse} for a $\sech$ envelope with various values of $\tau$. The time-domain waveforms were normalised to have amplitude 1\,Vm$^{-1}$.
When $\tau$ was larger, the multi-cycle THz pulse (red line in panel (a), $\tau=2$\,ps) tracked the oscillations of the carrier wave (dashed line), while the Fourier amplitude spectrum was relatively narrow and was centred at the carrier wave frequency (1\,THz). 
This waveform is representative of narrow-band THz pulses containing a few cycles, such as produced by some optical rectification schemes \cite{Mosley2023}, while mode-locked THz QCLs \cite{Barbieri2011} and THz FELs \cite{Green2016} generate multi-cycle THz pulses with even larger pulse envelopes.

When the envelope pulse duration was lowered to $\tau=200$\,fs or 500\,fs, the waveform and spectra (yellow and green lines in Fig.\ \ref{fig:sech-tau}) became more representative of broadband THz generation, such as achieved via typical optical rectification using laser amplifiers (e.g.\ in LiNbO$_3$ via the tilted-pulse method, or using ZnTe or GaP). 
For pulse durations below $\tau=100$\,fs, the spectrum became ultra-broadband: the spectral peak shifted to higher frequencies, and the gradient of the amplitude spectrum at high frequencies -- the \emph{spectral slope} -- became more shallow.
Noting that for a sech envelope the gradient of the amplitude spectrum on a logarithmic $y$-axis is linear at high frequencies $\omega \gg \omega_c$ (Figure \ref{fig:sech-tau}(b) and Figure \ref{fig:sech-v-gaussian}(b)), we derive an expression for the spectral slope.
In this limit, the amplitude spectrum can be calculated from Equation \ref{eq:CFT-THz-pulse}:
\begin{eqnarray}
  |\widetilde{E}(\omega\rightarrow \infty)| &\approx& |A| \frac{\pi \tau}{2} 
 \sech \left( \frac{\pi \tau}{2}  \omega \right) \nonumber \\ 
 &=& |A| \frac{\pi \tau}{2} \frac{2}{e^{\pi \tau \omega/2} + e^{-\pi \tau \omega/2}} \nonumber \\ &\approx& |A| \pi \tau e^{-\pi \tau \omega/2}. \label{eq:amplitude-limit}
\end{eqnarray}

\noindent From this equation, we can deduce that the amplitude spectrum drops exponentially at high frequencies for a sech pulse envelope, and hence has a negative straight line when plotted with a log $y$-axis (\eg Figure \ref{fig:sech-tau} (b)). 
Since $\ln |\widetilde{E}(\omega\rightarrow \infty)|= \ln (|A| \pi \tau) - \pi \tau \omega /2$, the gradient is $d|\widetilde{E}(\omega\rightarrow \infty)|/d\omega=-\pi \tau /2.$ 
This depends only on the duration of the pulse envelope. 
Hence the gradient of the linear spectral slope (on a logarithmic $y$-axis) can yield the duration $\tau$ of the THz pulse's envelope, assuming it has a sech shape. Similarly, if the THz pulse envelope was instead proportional to sech$^2(t/\tau)$, the high frequency limit of the Fourier spectral amplitude (Eq.\ \ref{eq:FTsech2}) is again exponential and has the same spectral slope.

\subsection{Influence of pulse arrival and phase offset}
\label{sec:THz-CEP}
%A more complete analytical expression for the THz pulse spectrum (including the pulse arrival time and carrier-envelope phase) can be obtained by taking the CFT of Equation \ref{eq:THz-pulse-time-domain},  yielding:
%
%\begin{flalign}
%\tcbhighmath{
%\begin{aligned}
% \widetilde{E}(\omega) = A \frac{\pi \tau}{2} e^{-i\omega t_e} 
% \left(  i  \cos(\omega_c \tCEP)  \right. & \left. \left[ \sech \left( \frac{\pi \tau}{2} \left( \omega-\omega_c \right) \right) -  \sech \left( \frac{\pi \tau}{2} \left( \omega+\omega_c \right) \right) \right] \right.  \\ 
% -\sin(\omega_c \tCEP ) & \left. \left[ \sech \left( \frac{\pi \tau}{2} \left( \omega-\omega_c \right) \right) +  \sech \left( \frac{\pi \tau}{2} \left( \omega+\omega_c \right) \right) \right]   \right).
%\end{aligned}
%}
%\end{flalign}
 
\begin{figure}[bt]
	\centering
	\includegraphics[width=1.0\columnwidth]{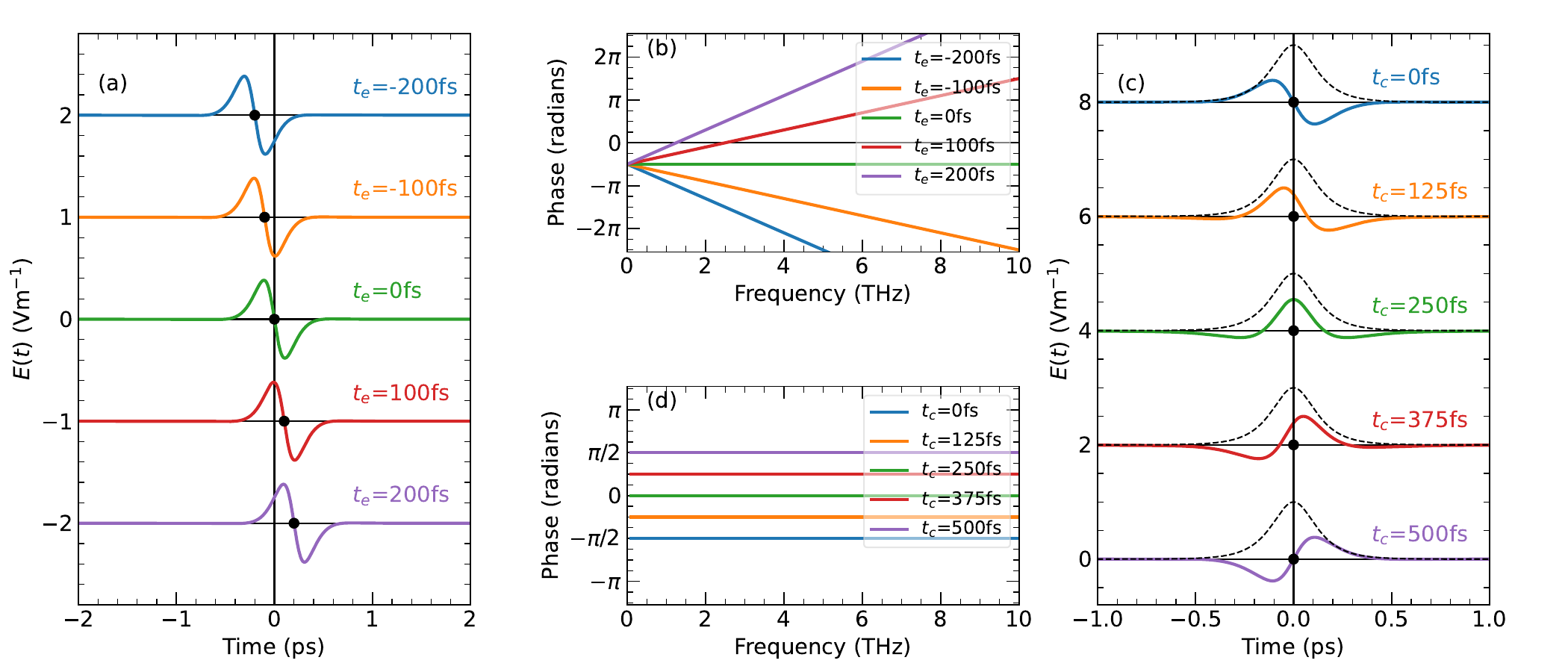}
	\caption{Influence of pulse arrival time and carrier-envelope phase. (a) THz time-domain pulses calculated using Eqn.\ \ref{eq:THz-pulse-time-domain} with $\omega/2\pi=1$\,THz, $\tau=100$\,fs, $\tCEP=0$ and various values of $t_e$. Black points indicate the pulse arrival time (mean arrival time method). Pulses are plotted offset vertically for clarity. (b) Unwrapped phase spectra from DFT of data in (a). (c) THz pulses with $t_e=0$, and varying carrier times $0\leq t_c \leq 500$\,fs, corresponding to different carrier-envelope phases $0\leq \phi_0 \leq \pi$. The dashed lines indicate the sech envelope, and the black points are the mean arrival time. (d) Phase spectra corresponding to the cases shown in (c).}
	\label{fig:t0}       
\end{figure}

%
The influence of the pulse arrival time and carrier-envelope phase is reported in Fig.\ \ref{fig:t0}.
In panels (a) and (b), single-cycle THz pulses are shown arriving at different times $t_e$ from -200\,fs to 200\,fs, and keeping $t_{\mathrm{CEP}}=0$ ($t_e=t_c$) to first discuss the case with zero carrier-envelope phase.
The amplitude spectra (not shown) are identical, but by the Fourier shift theorem changing the centre times of the THz pulse envelope (black dots in panel (a)) produces a linear gradient to the phase, shown unwrapped in panel (b).
The case with $t_e=0$ produces a constant absolute phase $-\pi/2$.

In order to produce physically realistic broadband time-domain pulses even in the case of finite carrier-envelope phase (as discussed above with reference to Ref.\ \cite{Ahmed2014}) we follow Ahmed {\emph{et al.}'s procedure and take the DFT of Eqn.\ \ref{eq:THz-pulse-time-domain} with $\phi_0=0$, multiplying the complex spectrum by the Heaviside step function, $\Theta(\omega)$, and taking the inverse DFT. 
This produces an analytic representation of the electric field, $\widetilde{E}(t)$, a complex time-domain function with no negative frequency components. 
The case with finite CEP can be correctly included by additionally multiplying the complex spectrum by a global phase term, $e^{i\phi_0}$, before the inverse DFT. The physical electric field is thus:
\begin{equation}
 E(t) = \Re \left( \mathcal{F}^{-1}\left[2 e^{i\phi_0}  \Theta(\omega) \widetilde{E}(\omega) \right]\right),
 \label{eq:analyticalRepTHzPulse}
\end{equation}

\noindent where $\phi_0=\omega_c t_{\mathrm{CEP}}=\omega_c (t_c-t_e)$ is the carrier-envelope phase shift. The factor of 2 ensures the same amplitude as the original electric field (since half of the amplitude is discarded by the Heaviside function).

Representative THz pulses for different carrier-envelope phases are reported in Fig.\ \ref{fig:t0}(c)-(d), obtained by setting $t_e=0$ and varying $t_c$.
As $t_c$ increased, the waveform changed shape from single-cycle ($t_c=0$) to become closer to a symmetric half cycle ($t_c=250$\,fs), as the carrier wave became more like a cosine.
A THz pulse with a symmetric shape about the pulse centre 
($\omega_c t_c = \pm \pi/2$) is an even function ($E(-t)=E(t)$), and hence via Fourier theory has a purely real Fourier spectrum with zero phase.
Conversely, $\omega_c t_c=\{ 0, \pi\}$ results in an odd function ($E(-t)=-E(t)$) and a purely imaginary spectrum with $\phi (\omega )=\pm \pi/2$.
These cases can be seen in panel (d), where $t_c=250$\,fs corresponds to $\omega_c t_c=\pi/2$, and $t_c=500$\,fs corresponds to $\omega_c t_c=\pi$. 
%Intermediate carrier phases ($t_c =100$\,fs or 400\,fs) result in more complex phase spectra. 

%\begin{figure}[tb]
%	\centering
%	\includegraphics[width=1.0\columnwidth]{figure-CEP.pdf}
%	\caption{THz time-domain pulses for $\omega_c/2\pi=1$\,THz and $\tau=100$\,fs, with envelope time $t_e=0$, and different carrier times in the range $0\leq t_c \leq 500$\,fs, corresponding to carrier phase $0\leq \omega_c t_c \leq \pi$. Pulses are plotted offset vertically for clarity. The dots indicates the calculated $t_{g}$. (b) and (c) show the amplitude and phase spectra corresponding to the cases shown in (a), calculated from the DFT (solid lines) or using the analytical CFT result (dashed lines).}
%	\label{fig:CEP}       
%\end{figure}

%For few-cycle pulses, the amplitude spectrum also changes in shape with the carrier phase (Figure \ref{fig:t0}(d)): when the phase is such that the pulse is more like a half-cycle, the spectrum gains amplitude towards low frequency.
%The spectral slope at high frequency, however, is independent of the carrier phase.

\subsection{Pulse arrival time}

The definition of the phase-accurate DFT in Sec.\ \ref{sec:dft} implicitly assumes that the time, $T_0$, from the first point in the series to the centre of the pulse, is known. 
While this is the case if the waveform is calculated from an analytical expression (\emph{e.g.}\ Eqn.\ \ref{eq:THz-pulse-time-domain}), for an experimental waveform the value of $T_0$ needs to be input or determined by the DFT routine. 
Here we provide various approaches to determining the pulse arrival time:

\begin{enumerate}
 \item \textbf{Educated guess}. If the THz pulse is close to single-cycle in shape, manually read the time delay where the electric field changes sign between the positive and negative extrema. If the THz pulse is more like a symmetric half-cycle, take the time delay of the largest peak. 
 \item \textbf{Mean arrival time}. Calculate the mean arrival time of the pulse, based on the first-order moment of the normalised intensity, $\left< t \right>=\int_{-\infty}^{\infty} t I(t) dt$, where $I(t)=E^2(t)/\int_{-\infty}^{\infty}E^2(t)dt$.
 \item \textbf{Time-domain fit}. Fit an analytical model of the electric field, such as Eqn.\ \ref{eq:THz-pulse-time-domain}, to determine the arrival time of the envelope. 
 \item \textbf{Group delay}. The first derivative of the spectral phase $\phi(\omega)$ is called the group delay, $t_g(\omega)= {d\phi}/{d\omega}$. Evaluating the group delay at the frequency of maximum spectral amplitude, $\omega_m$, allows the arrival time of the pulse to be determined. This result can be understood from the Fourier shift theorem, as the linear component of the spectral phase corresponds to a delay in the time domain.

\end{enumerate}

% written using a Taylor series expansion about the frequency of maximum spectral amplitude, $\omega_m$, as
%\begin{equation}
% \phi(\omega) = \phi_0 +  \frac{(\omega-\omega_m)}{1!} \left. \frac{d\phi}{d\omega}\right|_{\omega=\omega_m} +  \frac{(\omega-\omega_m)^2}{2!} \left. \frac{d^2\phi}{d\omega^2}\right|_{\omega=\omega_m} + ...
%\end{equation}
%\noindent The first order derivative of $\phi(\omega)$ is called the group delay, $t_g(\omega)= {d\phi}/{d\omega}$. 
%Evaluating the group delay at the maximum frequency cancels out all the terms apart from the first derivative of the phase: $t_g(\omega_c)= \left. {d\phi}/{d\omega}\right|_{\omega=\omega_m}$, yielding the arrival time of the centre of the pulse envelope. 

The solid circles in Fig.\ \ref{fig:t0}(a) and (c) indicate the pulse arrival times calculated using the mean arrival time method, and agree exactly with the input envelope times, indicating that the method works in this case. 
The ``educated guess'' method requires no extra calculations but requires the experimenter's input, and is thus not automated or as precise as the other methods. 
With finite carrier-envelope phase (e.g.\ $t_c=125$\,fs in Fig.\ 4(c)), the educated guess method predicts the wrong arrival time, since the zero crossing is not at the peak of the pulse envelope.
The mean arrival time can be calculated automatically and rapidly via numerical integration, and is accurate for pulses that are not too complex.  
For example, it can be used to correct for arrival time differences between two THz pulses in dual-beam THz spectroscopy \cite{Chopra2025}.
However it cannot be used for more complex waveforms: for instance if the data set comprises a pulse train created by internal reflections, the time returned by the integral will be pulled later in time by the pulse echoes.
The time-domain fit method is more complex to implement, but can be extended to cope with situations involving multiple pulses.
Finally, the group-delay method is simpler than the time-domain fit to implement, and yields the same results as the mean arrival time method if the pulse is relatively simple (i.e.\ if the spectral phase is relatively flat). 
However the group-delay method stumbles if the spectral phase has a complex shape at the frequency of the maximum spectral amplitude. 
This can be the case, for example, if the THz pulse has undergone strong dispersion as a result of a resonance near the maximum frequency, or from multiple internal reflections \cite{Knorr2018}.

%If the pulse has a linear phase spectrum (zero chirp), as pictured in Fig.\ \ref{fig:t0}(b), then the Taylor series tells us that the second order terms (and higher orders) are zero. 
%Thus a change in the phase proportional to the frequency only shifts the pulse in time, while higher order terms can alter the chirp (variations in the arrival time with angular frequency).
%The second order term in the Taylor series contains the group delay dispersion, $\mathrm{GDD}=\frac{d t_g}{d\omega}=\frac{d^2\phi}{d\omega^2}$, which is the leading term for the chirp.

%\subsection{Chirp}

%Using this expression yields the arrival time of the pulse envelope, shown by the circles in Figs.\ \ref{fig:CEP} and \ref{fig:t0}.
%The group delay dispersion is $$\frac{d t_g}{d\omega}=\frac{d^2 \phi}{d \omega^2}.$$

\section{Experimental results}
\label{sec:exp}

Armed with the insights from the previous sections, we present in Sec.\ \ref{sec:exp-amplitude} an analysis of experimental pulses from typical broadband and narrowband THz sources: two laser-based THz sources (optical rectification in LiNbO$_3$ and a spintronic emitter) and a THz free-electron laser.
The spectral energy density allows a direct comparison of the relative performance of each source, while fits using the phenomenological model from Section \ref{sec:THz} provide insights into pulse parameters.
We then discuss, in Section \ref{sec:exp-literature}, literature reports of the spectral phase of THz and mid-IR sources obtained using electro-optic sampling, before providing a step-by-step analysis of the spectral phase of a laser oscillator-based THz system in Sec.\ \ref{sec:exp-phase}. 

\subsection{Spectral energy density}
\label{sec:exp-amplitude}

Using the DFT with correct amplitude scaling (Sec.\ \ref{sec:dft}, Eq.\ \ref{eq:DFT5}) allows the spectral energy density as a function of angular frequency or frequency, $u(\omega)$ or $u(f)$, to be determined experimentally.
Our aim in this section is not to claim the highest time-domain $E(t)$, pulse energy or spectral energy density: rather the aim is to show how the spectral energy density of sources of THz pulses can be quantitatively compared.

Starting from the time-domain definition of the instantaneous energy density of an electromagnetic wave in free space, $u(t)=\epsilon_0 c |E(t)|^2$, the energy per pulse per unit area, $U$, can be calculated from 
\begin{eqnarray}
U=\int_{-\infty}^{\infty} u(t) dt &=& \epsilon_0 c \int_{-\infty}^{\infty}  |E(t)|^2 dt \nonumber \\ 
&=& \frac{\epsilon_0 c}{2 \pi} \int_{-\infty}^{\infty} |E(\omega)|^2 d\omega = \int_{-\infty}^{\infty}  u(\omega) d\omega,\label{eq:spectral}
\end{eqnarray}

\noindent where Parseval's theorem (Eq.\ \ref{eq:Parseval}) was used to convert the time-domain integral to a frequency domain integral. 
Therefore the spectral energy density (assuming our `Physics' definition of the scaling in the Fourier transform) is $u(\omega) = \epsilon_0 c |\widetilde{E}(\omega)|^2/(2 \pi)$ or $u(f) = \epsilon_0 c |\widetilde{E}(f)|^2$.
The SI units for $u(f)$ are J m$^{-2}$ s, or J m$^{-2}$ Hz$^{-1}$, and hence plotting the spectral energy density in units of J m$^{-2}$ THz$^{-1}$ allows an easy estimate of the energy fluence, $U$. 
For example, a THz source with 1\,THz bandwidth and $u(f)\simeq 1$\,J m$^{-2}$ THz$^{-1}$ will have $U\simeq 1$\,J m$^{-2}$. 

If the spatial profile of the THz beam is known, for example using a THz camera or a knife-edge measurement, then the total energy per pulse, $E_{\mathrm{pulse}}$, can be obtained by integrating over the area:
\begin{equation}
 E_{\mathrm{pulse}} = U \int \int g(x,y) dx dy = 2\pi \sigma^2 U \label{eq:Upulse},
\end{equation}
\noindent where $g(x,y)$ is the beam profile, assumed here to be a 2D Gaussian with standard deviation $\sigma$ in both cross-sectional co-ordinates, $x$ and $y$.
Here $U$ is the peak energy per pulse per unit area at the centre of the THz beam, which is the area sampled in electro-optic sampling with a small gate beam.
In the example of $U\simeq 1$\,J m$^{-2}$, a reasonably small THz beam (say $\sigma=400$\,$\mu$m) thus has $E_{\mathrm{pulse}}\simeq 1$\,$\mu$J.
The average power of the THz beam is then the repetition rate of the source multiplied by $E_{\mathrm{pulse}}$.
If the amplitude of the time-domain electric field $E(t)$ is not determined directly in a THz spectrometer (e.g.\ the electro-optic or photoconductive signal is in arbitrary units), then common practise is to use the measured THz pulse energy and beam profile to deduce the electric field strength by substituting Eq.\ \ref{eq:spectral} in Eq.\ \ref{eq:Upulse} and scaling the electric field until the calculated pulse energy matches the experiment.

In Figure \ref{fig:experiment-amplitude} we show the THz electric field of THz pulses from three different sources: (a) the TELBE superradiant undulator source \cite{Chen2022}, (b) a laser-driven source using LiNbO$_3$, and (c) a laser-excited spintronic emitter.
The spectral energy density, $u(f)$, of each source can be readily compared in Fig.\ \ref{fig:experiment-amplitude}(d). 
The TELBE source produced a THz beam with centre frequency 0.68\,THz, repetition rate $R=50$\,kHz and an average power of around 32\,mW, corresponding to $E_{\mathrm{pulse}}=0.65$\,$\mu$J.
The multi-cycle THz pulse was sampled by electro-optic detection using a synchronised slave laser \cite{Chen2022}, and the THz electric field is reported in Fig.\ \ref{fig:experiment-amplitude}(a).
The time-domain electric field amplitude was scaled such that the integrated pulse energy matched the experimental value, and assuming the pulse was focusable to $\sigma=300$\,$\mu$m.
The peak time-domain electric field is then around $87$\,kV/cm, or a maximum spectral energy density $|u(f)|= 13.5$\,Jm$^{-2}$THz$^{-1}$, Fig.\ \ref{fig:experiment-amplitude}(d).
The mean time of the pulse was 4.0\,ps past the maximum electric field, owing to the asymmetry of the pulse. 
As a class of intense narrowband sources with high spectral energy density, THz FELs have found use recently in studies of THz harmonic generation \cite{Chen2022,Meng2023}. 
\begin{figure}[tb]
	\centering
	\includegraphics[width=1\columnwidth]{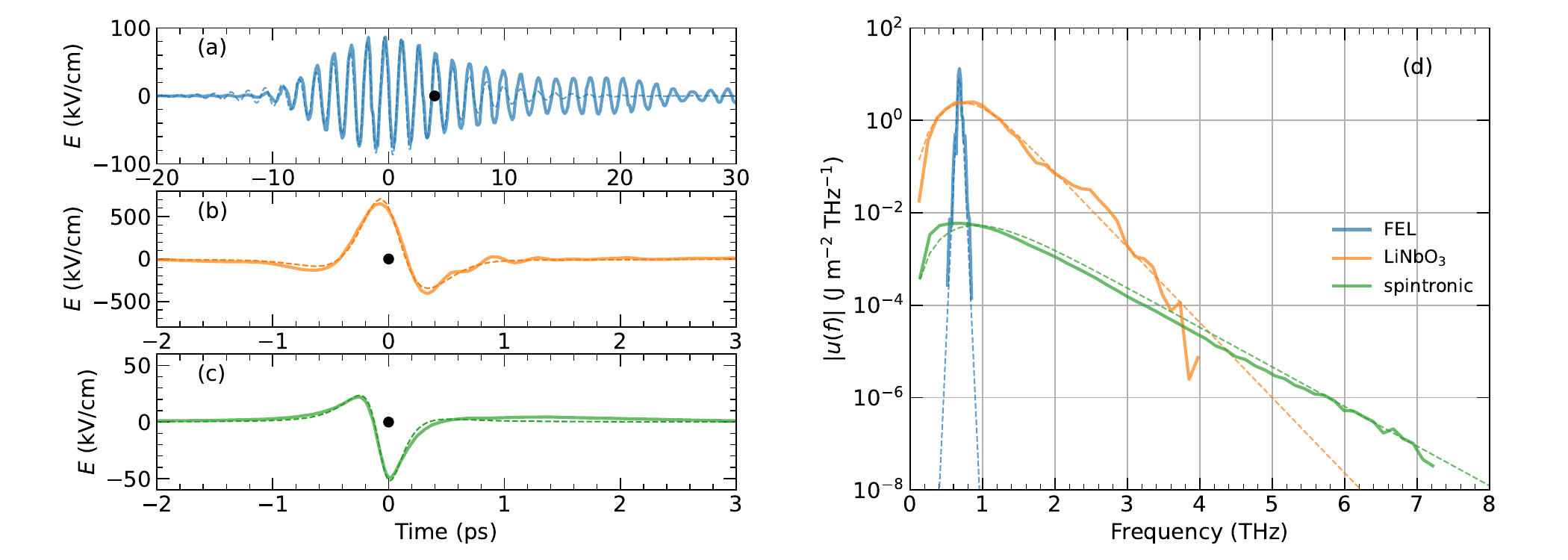}
	\caption{(a) Multi-cycle THz pulse produced by TELBE, running at 0.68\,THz. The black spot is the pulse arrival time calculated from the mean-time method. (b) Electric field produced by LiNbO$_3$ photoexcited by a 1\,mJ, 800\,nm, 40fs laser pulse; (c) As (b) but for a spintronic emitter. (d) Spectral energy density $u(f)$ for the THz sources in (a)-(c), shown over frequency range where the signal exceeded the noise floor (which varied from experiment to experiment).}
	\label{fig:experiment-amplitude}       
\end{figure}
 
The peak time-domain electric field produced following fs excitation of LiNbO$_3$ is reported in Fig.\ \ref{fig:experiment-amplitude}(b), following the tilted-pulse front method \cite{Hirori2011}, using a 40\,fs, 1\,kHz Ti:sapphire amplifier and a 1\,mJ excitation pulse, and electro-optic detection using a 200\,$\mu$m-thick GaP crystal \cite{Riccardo2025}. 
A peak electric field of around 650\,kV/cm and THz pulse energy  $E_{\mathrm{pulse}}=1$\,$\mu$J were obtained, with a beam radius $\sigma=200$\,$\mu$m measured by a THz camera.
While the energy per pulse exceeded that for the TELBE data, the peak spectral energy density was around 6 times lower, $u(f)=2.5$\,Jm$^{-2}$THz$^{-1}$, as a result of the more broadband spectrum, reaching 3.8\,THz before hitting the noise floor.
The same fs laser and EOS scheme was instead used to determine the THz pulse from a commercial trilayer spintronic emitter \cite{Seifert2016} from TeraSpinTec, in an optical-pump, THz probe spectrometer \cite{Ren2023} that achieved around 50\,kV/cm peak electric field ($\sigma=200$\,$\mu$m, $E_{\mathrm{pulse}}=3.8$\,nJ), Fig.\ \ref{fig:experiment-amplitude}(c).
Below 3.8\,THz, the spectral energy density was lower than that for the LiNbO$_3$ source, but the more broadband nature of the emission allowed greater energy density at higher frequencies.
Fits of the phenomenological time-domain model reported in Section 3 (Equation \ref{eq:analyticalRepTHzPulse}, using a sech pulse) are also shown in the figure (dashed lines), with THz pulse durations $\tau=100$\,fs, $\tau=200$\,fs and $\tau=4$\,ps for the spintronic source, LiNbO$_3$ and TELBE data, respectively. 
Note that the fit quality in the time-domain was reasonable for the laser-based THz sources, but the fit did not capture the more complex pulse structure of TELBE at times above 5\,ps.

\subsection{Spectral phase: literature reports}
\label{sec:exp-literature}
Only a limited number of papers in the time-domain spectroscopy community have reported the spectral phase, of which the following is a representative (but non-exhaustive!) list. 
Leitenstorfer \emph{et al.}\ \cite{Leitenstorfer1999} reported ultrabroadband amplitude and phase spectra from $p-i-n$ diodes, with complex phase that decreased from 3\,rad at low frequency to 1\,rad at 3\,THz, before remaining relatively flat up to the LO-phonon frequency of InP, where it varied rapidly.
K\"{u}bler \emph{et al}.\ showed the electric field of ultrabroadband mid-infrared pulses produced and detected by GaSe, with a spectral maximum at 30\,THz \cite{Kubler2004}. 
They assumed a Gaussian envelope for the 3-cycle electric field pulse, which had 28\,fs duration.
The spectral phase was below 1\,rad from 10\,THz to 100\,THz, which is consistent with their cosine-like electro-optic signal and our discussion above with reference to Fig.\ \ref{fig:t0}.
Knorr \emph{et al}.\ reported the electric field of 20\,THz mid-infrared pulses detected using thin GaSe flakes \cite{Knorr2018}, and showed a relatively flat spectral phase from 15\,THz to 30\,THz, but with a rapidly-increasing spectral phase at lower frequencies.
The carrier-envelope phase was not determined, as the time-domain peak electric field was assumed to be the centre of the pulse envelope.
Ultrabroadband THz pulses from spintronic emitters were first reported by Seifert \emph{et al}.\ \cite{Seifert2016}, and were found to have close to zero spectral phase over a 30\,THz bandwidth, signifying chirp-free (Fourier limited) pulse durations. 
However, this cannot be seen straightforwardly from the reported time-domain electro-optic signal, which had a finite pulse arrival time (250\,fs), which would imply a linear slope to the phase (not reported), and an asymmetric pulse shape, which should give finite (non-zero) CEP (Fig.\ \ref{fig:t0}).
%Understanding the exact pulse envelope and spectral phase of different THz emission processes may well yield insights into the underlying physical mechanisms.

In the above experimental works the spectral phase was not used to uncover the carrier-envelope phase or the group delay. 
An extension of frequency-resolved optical grating (FROG) was reported to yield the CEP of a half-cycle THz pulse \cite{Snedden2015} to be around 2\,$^{\circ}$.
While the electric-field waveform from the retrieved FROG spectrogram was similar to that directly obtained from EOS, the spectral phase from EOS was not reported.
Control of the CEP for a THz pulse was reported using Fresnel prisms \cite{Kawada2016}, where the input pulses had a complicated $\phi(\omega)$ ranging from $\phi(0.5$THz$)=-1.5$\,rad, to $\phi(1.75$THz$)=0$\,rad, to $\phi(2.5$THz$)=-2.5$\,rad. 
Although the origin of this complex $\phi(\omega)$ was not discussed, and the absolute spectral phase was not used to determine the CEP, rotating the Fresnel prisms did produce the expected changes in CEP (using the difference in spectral phase for different prism angles).
Allerbeck \emph{et al.} demonstrated how to modify the CEP of a THz pulse using frustrated total internal reflection in a prism coupled to a metal mirror \cite{Allerbeck2023}. 
A carrier-envelope phase model (similar to Eqn.\ \ref{eq:THz-pulse-time-domain} but with a Gaussian envelope), and which is not fully valid for single-cycle pulses with arbitrary CEP \cite{Ahmed2014}, was used to fit the time-domain waveforms and infer the CEP \cite{Allerbeck2023}. 
Finally, Zhang \emph{et al}.\ examined the spectral phase of the THz radiation produced by a DC-biased laser-induced plasma in air, and found that changing the position of the bias along the filament altered the CEP of the THz pulse, as deduced from the zero-frequency limit of the spectral phase \cite{Zhang2020}.

\subsection{Experimental phase spectra}
\label{sec:exp-phase}
Here we provide a detailed analysis of the absolute spectral phase of a typical source of pulsed THz radiation, in order to demonstrate that the CEP, group delay and chirp can be readily obtained from $\phi(\omega)$.  
We recorded the time-domain THz waveform from an interdigitated photoconductive switch, made on semi-insulating GaAs and photoexcited with 800\,nm central wavelength pulses from a mode-locked Ti:sapphire oscillator (Newport Spectra Physics Mai Tai) with a pulse duration of 100\,fs \cite{Chopra2023a}.  
Four off-axis parabolic mirrors (3'' focal length, 2'' diameter) were placed in a $(a,a)(a,a)$ geometry, as described in Ref.\ \cite{Chopra2023}, to collect and focus the THz radiation to the sample plane and then the detection crystal.
Electro-optic sampling was performed using a 200\,$\mu$m-thick $(111)$-oriented GaP detection crystal on a 1\,mm-thick $(100)$ GaP substrate.
The entire THz beam path was purged with nitrogen gas.
\begin{figure}[tb]
	\centering
	\includegraphics[width=1.0\columnwidth]{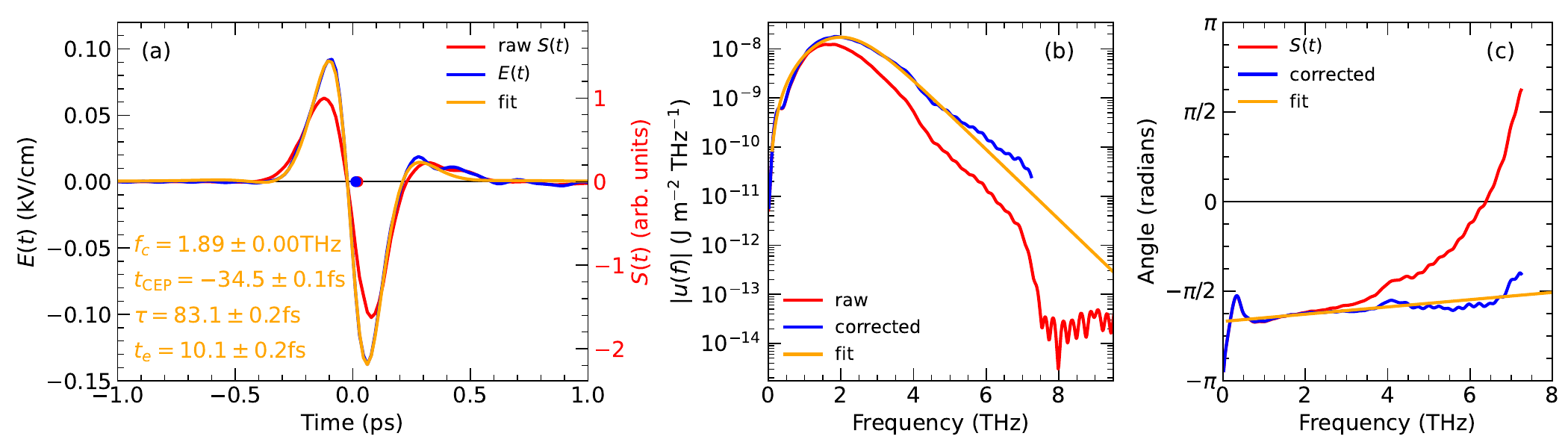}
	\caption{Experimental THz pulse emitted by a photoconductive emitter, compared with the THz pulse model. (a) Raw electro-optic signal $S(t)$ before deconvolving the detector's response (red); after deconvolving the detector's response (blue). A fit using Equation \ref{eq:THz-pulse-time-domain} is shown by the orange line. (b) and (c) show the amplitude and phase DFT spectra of the data in (a).}
	\label{fig:experiment}       
\end{figure}

We report in Figure \ref{fig:experiment} the time- and frequency-domain electro-optic signals, $S(t)$ and $\widetilde{S}(\omega)$, obtained for this spectrometer (red curves). 
The pulse was close to single-cycle in shape [panel (a)], and had frequency components spanning from 100\,GHz to 7.5\,THz [panel (b)] before hitting the noise floor.
The absolute phase of the electro-optic signal, reported in Fig.\ \ref{fig:experiment}(c), started close to $-\pi/2$ at low frequencies (red line), as expected based on the discussion in Section \ref{sec:THz}. 
While the spectral phase was relatively flat to 3.5\,THz, at higher frequencies the phase began to increase dramatically.
This can be understood as a result of the electro-optic sampling process: the measured electro-optic signal does not have exactly the same temporal shape as the true electric field of the THz pulse incident on the detection crystal \cite{Gallot1999,Leitenstorfer1999}.
For example, in the crystal the phase velocity of THz radiation and the group velocity of the infrared gate pulse are different, and hence after propagating some distance through the crystal they may no longer be in phase.
This introduces a non-linear spectral phase, or chirp, to the electro-optic signal.
Deconvolving the detector's complex response function is well described in the literature \cite{Gallot1999,Leitenstorfer1999,Casalbuoni2008}, and we detail our particular implementation in Appendix \ref{app:EOS}.

Correcting for the response function, $\widetilde{R}(\omega)$, of the 200\,$\mu$m-thick GaP electro-optic crystal over the range below 7.5\,THz produces the blue lines in panels (b) and (c), which show the amplitude and phase of the electric field.
Above 7.5\,THz the electro-optic signal could not be corrected, as the response function had a minimum strong enough that the electro-optic signal dropped below the noise floor.
The spectral amplitude of the electric field is marginally higher than the electro-optic signal in the range 2\,THz to 7.5\,THz as a result of $|\widetilde{R}(\omega)|$ being less than 1 in this range.
After correction, the spectral phase is much flatter, implying that the THz pulse incident onto the detector was relatively free of chirp.

The time-domain pulse, $E(t)$, was close to single-cycle in shape, and can be accurately modelled via Equation \ref{eq:analyticalRepTHzPulse}, as illustrated by the yellow line in Fig.\ \ref{fig:experiment}(a).
A time-domain fit yielded a pulse arrival time $t_e=10.1\pm0.2$\,fs and duration $\tau=83\pm1$\,fs, along with a carrier frequency $f_c=1.89\pm0.01$\,THz and carrier-envelope time $\tCEP=-35\pm1$\,fs.
The amplitude and phase of the model in the frequency domain are also in good agreement with the experimental data.
In particular, the small negative $\tCEP$ reduced the constant spectral phase $\phi(\omega\rightarrow 0)$ below $-\pi/2$ (panel (c)), while the small $t_e$ gave the spectral phase a small positive slope.
It is also clear that the THz pulse envelope was as short as possible: it had minimal chirp, and a duration that was similar to the specified duration of the fs laser used (100\,fs). 

Finally, the experimental amplitude spectra from photoconductive emission (Fig.\ \ref{fig:experiment}), the spintronic emitter and optical rectification (Fig.\ \ref{fig:experiment-amplitude}) all exhibited an exponential decay at high frequency.
Fourier theory attributes the spectral behaviour at high frequencies to the fastest time-domain dynamics of the THz pulse, which should correspond to the fastest physical process, potentially the temporal intensity envelope, $I(t)\propto E^2(t)$, of the fs laser used (here $E(t)$ refers to the fs pulse's electric field).
In this situation, the THz spectral tail inherits its decay behaviour from the pump laser's envelope, for example a Gaussian spectrum from a Gaussian input time-domain pulse (Fig.\ \ref{fig:sech-v-gaussian}b). 
THz emission via optical rectification follows $E_{\mathrm{THz}}(t)\propto \frac{d^2}{dt^2}[P(t)] \propto \chi^{(2)} \frac{d^2}{dt^2} [E^2(t)] \propto \frac{d^2}{dt^2} [I(t)]$.
In the frequency domain, taking the Fourier transform yields $E_{\mathrm{THz}}(\omega)\propto \omega^2 I(\omega)$, showing that the THz spectrum follows the incident pulse's spectrum.
A similar argument can be made for photoconductive or spintronic sources, which are based on transient photocurrents $J(t)$: the far-field THz emission is $E_{\mathrm{THz}}(t)\propto dJ(t)/dt$ and hence the initial rise in photocurrent $J(t)\propto E^2(t)$ tracks the intensity envelope of the fs laser. 
If this is the fastest dynamic process it will control the shape of the THz spectrum at high frequencies.

The observed exponential tail in frequency is is consistent with a time-domain sech or sech$^2$ pulse envelope (see Eqs.\ \ref{eq:amplitude-limit} and \ref{eq:FTsech2}), which both have exponential tails at high frequencies. 
If the time-domain pulse had a gaussian envelope, the spectral amplitude would drop more rapidly than seen experimentally (Fig.\ \ref{fig:sech-v-gaussian}b). 
Sech$^2$ intensity envelopes are typical of fs laser oscillators and can thus lead to a sech$^2$ time-domain component in the emitted THz field, or an exponential frequency dependence. 
If a Gaussian spectrum or time-domain pulse envelope were instead obtained, it might imply a process has randomised the arrival time of the THz pulse somewhat (e.g.\ random error in a rapid-scan delay line), rather than a THz pulse with bandwidth limited by a sech-like fs pulse envelope.

%This can be understood from 
%\begin{equation}
% E_{\mathrm{THz}}(t) \propto \frac{dJ(t)}{dt} \propto \frac{dN(t)}{dt} \rightarrow E_{\mathrm{THz}}(\omega) \propto i \omega J(\omega)
%\end{equation}

\section{Conclusion}
In this Article we presented aspects of Fourier theory as applied to electromagnetic pulses, including how to ensure phase- and amplitude-accuracy when using the discrete Fourier transform (Section 2), and a simple model of single- and multi-cycle light pulses (Section 3).
Amplitude accuracy allowed the spectral energy density of disparate experimental THz sources to be compared (Section 4), while phase-accuracy allowed the absolute phase to be calculated robustly, yielding the carrier-envelope phase, pulse arrival time and inherent chirp of a THz pulse.
Our aim was to provide a single reference that provides insights into pulse analysis for time-domain ultrafast spectroscopy methods.
Although we presented and validated our methods using data from THz time-domain spectroscopy, the approaches summarised here are equally applicable to mid-infrared and visible time-domain spectroscopic methods that use electro-optic sampling (or similar approaches) to uncover the electric field of light pulses \cite{Huber2000,Zimin2021,Herbst2022}.
%The time-domain model and frequency-domain DFT analysis, when compared to experimental data, provides useful insights into the properties of such pulses, namely the pulse arrival time (group delay), carrier-envelope phase, and inherent chirp of the pulse. 
The accurate determination of the pulse arrival time may prove useful in THz time-of-flight, layer thickness or tomography applications \cite{Leitenstorfer2023}, and further helps to correct arrival time differences in dual-beam THz spectroscopy \cite{Chopra2025}. 
The ability to find the carrier-envelope phase easily from the absolute spectral phase will allow setups to be optimised for maximal peak electric field \cite{Allerbeck2023}, relevant for nonlinear THz spectroscopy techniques such as 2D THz and THz STM.
The methods reviewed may prove useful for researchers in other areas of ultrafast spectroscopy, such as multidimensional coherent spectroscopy \cite{Donaldson2022}, which use Fourier transforms to perform data analysis.

\appendix
\section{Useful results from continuous Fourier transforms}
\label{app:Fourier}

%The continuous Fourier transform (CFT) converts a continuous time-domain function to a frequency-domain function, allowing the relative amplitude and phase of different frequency components to be established.
%In THz-TDS it returns a complex electric field spectrum $\widetilde{E}(\omega)$ from a real-valued input, the time-domain THz electric field $E(t)$.
%The forward CFT, denoted by $\mathcal{F}$, can be defined as
%\begin{equation}
%\tcbhighmath[drop fuzzy shadow]
%{\mathcal{F}[E(t)]=\widetilde{E}(\omega) = \int_{-\infty}^{\infty} E(t) e^{i\omega t} \,dt.}
%\label{eq:forwardFT}
%\end{equation}

%The inverse Fourier transform, $\mathcal{F}^{-1}$, can return the electric field as a function of time from the spectrum, via 
%\begin{equation}
%\mathcal{F}^{-1}[\widetilde{E}(\omega)] = E(t) = \frac{1}{2\pi} \int_{-\infty}^{\infty} \widetilde{E}(\omega) e^{-i\omega t} \,d\omega.
%\label{eq:inverseFT}
%\end{equation}

Some standard Fourier transforms that are of use to model ultrafast pulses of light are as follows.

\emph{Sine.} The strongest frequency component of a light pulse is its ``carrier frequency,'' the frequency of the maximum spectral amplitude.
For a time-domain sine wave $f(t)=\sin(\omega_c t)$ oscillating with angular carrier frequency $\omega_c$ the CFT is 
\begin{equation} 
\widetilde{f}(\omega)=i\pi \left( \delta(\omega-\omega_c) - \delta(\omega + \omega_c) \right).
\label{eq:FTsine}
\end{equation}

\emph{Gaussian.} As a standard function, the Gaussian or normal distribution is often used to model the envelope of a light pulse. 
For a Gaussian profile centred at time $t=0$, the pulse can be described by $g(t)=A e^{-t^2/2\sigma^2}$, where a shorter pulse has smaller standard deviation $\sigma$.
The CFT of $g(t)$ is then 
\begin{equation}
\widetilde{g}(\omega)=\int_{-\infty}^{\infty}g(t) e^{i\omega t}dt = \sqrt{2 \pi \sigma^2} e^{-\omega^2 \sigma^2 /2}, 
 \label{eq:FTgaussian}
\end{equation}

\noindent and hence the Fourier transform of a Gaussian time-domain signal has a Gaussian profile in the frequency domain.

\emph{sech.} The hyperbolic secant function, $\sech(t)=1/\cosh(t)=2/(e^t+e^{-t})$, is often used to model the pulse envelope of light pulses from femtosecond lasers. 
Writing the pulse envelope as $f_e(t)=\sech(t/\tau)$ for pulse duration $\tau$, the CFT is 
\begin{equation}
\widetilde{f}_e(\omega)=\pi \tau \sech\left( \frac{\pi \tau}{2} \omega \right),
\label{eq:FTsech}
\end{equation}

\noindent and hence the Fourier transform of a $\sech$ time-domain signal also has a $\sech$ profile in the frequency domain.

\emph{sech$^2$.} Since the intensity envelopes of femtosecond laser pulses are often modelled as being proportional to $\sech^2(t/\tau)$, it is useful to know the corresponding CFT:
\begin{equation}
\mathcal{F}[\sech^2(t/\tau)]=\frac{\pi \tau}{2} 
\frac{\omega \tau}{\sinh\left( \frac{\pi \tau \omega}{2} \right)} = \frac{\pi \tau^2 \omega}{e^{\pi \tau \omega/2} - e^{-\pi \tau \omega/2}},
\label{eq:FTsech2}
\end{equation}

\noindent where the last expression can be readily taken in the high frequency limit to show that the spectrum decays exponentially at high frequencies.

\emph{Parseval's theorem}. The integral of the modulus squared of a function in the time-domain and frequency-domain are linked by Parseval's theorem:
\begin{equation}                                                            
\int_{-\infty}^{\infty}|E(t)|^2 dt = \frac{1}{2\pi} \int_{-\infty}^{\infty} |\widetilde{E}(\omega)|^2 d\omega,
\label{eq:Parseval}
\end{equation}

\noindent or in discrete form:
\begin{equation}   
\sum_{m=-\infty}^{\infty}|E_m(t_m)|^2 \delta t = \frac{1}{2\pi} \sum_{k=-\infty}^{\infty} |\widetilde{E}_k(\omega_k)|^2 \delta \omega_k.
\label{eq:ParsevalDFT}
\end{equation}

\noindent for discrete times $t_m$ and discrete angular frequencies $\delta \omega_k$. 
Parseval's theorem can be used to verify numerically whether the amplitude spectrum of the discrete Fourier transform is scaled correctly, by comparison to the sum (or integral) of the time-domain.

\emph{Fourier shift theorem}. For a function $f(t)$ with Fourier transform $\widetilde{f}(\omega)$, if the function is shifted in the time-domain to $f(t-T_0)$, then the Fourier transform of the shifted function can be written 
\begin{equation}
\mathcal{F}[f(t - T_0)] = e^{i\omega T_0} \widetilde{f}(\omega).
\label{eq:Fourier time-shift}
\end{equation}
Thus the Fourier transform of the time-shifted function has had its linear phase changed by $\omega T_0$ relative to the unshifted Fourier transform.

\emph{Product with sine}. For a time-domain function $h(t)=f(t) \sin(\omega_c t)$, its CFT can be shown to be 
\begin{equation}                                                            
\widetilde{h}(\omega)=\frac{\widetilde{f}(\omega+\omega_c)-\widetilde{f}(\omega-\omega_c)}{2i},
\label{eq:product with sine}
\end{equation}
 
\noindent allowing $\widetilde{h}(\omega)$ to be determined if $\widetilde{f}(\omega)$ is already known.

%\emph{Convolution theorem}. If a time-domain function $h(t)$ is the convolution of two functions, $f(t)$ and $g(t)$, written $h(t) = f(t) \circledast g(t) = \int_{-\infty}^{\infty} f(t) g(t-u) du$, then the Fourier transform of $h(t)$ is
%$$\widetilde{h}(\omega)=\mathcal{F}[h(t)]=\mathcal{F}[f(t)\circledast g(t)]= \widetilde{f}(\omega) \widetilde{g}(\omega).$$
%Conversely, if a time-domain function is the product of two other time-domain functions, \ie $h(t)=f(t)g(t)$, then in the frequency domain the Fourier spectrum is linked to the convolution of the Fourier transforms of $f(t)$ and $g(t)$ by 
%$$\widetilde{h}(\omega)=\frac{1}{2\pi} \widetilde{f}(\omega)\circledast \widetilde{g}(\omega).$$

%\include{scrap}

\section{Discrete Fourier Transform}
\label{app:DFT}
To derive the phase-corrected DFT of Eqn.\ \ref{eq:DFT5} we return to the example of the discretely sampled Gaussian.
This peaks at element $m_0$ (time $T_0$), and hence has the form $g_m=A e^{-(t_m-T_0)^2}/2\sigma^2$. 
Using Equation \ref{eq:DFT3} the DFT of the discretely-sampled Gaussian, $\widetilde{g}_k$, therefore approximates the continuous Fourier transform of the time-shifted continuous function $g(t-T_0)$, or:
\begin{equation}
\widetilde{g}_k \simeq \mathcal{F}[g(t-T_0)] = e^{i\omega t_0} \widetilde{g}(\omega)\label{eq:dft-result}
\end{equation}
\noindent where $\widetilde{g}(\omega)$ is as above, and the second equality is via the Fourier time-shift theorem.
The CFT of the time-shifted pulse, and the DFT of the Gaussian series, therefore both have a finite argument, $\omega T_0$, rather than the zero phase expected for the CFT of a Gaussian centred at $t=0$.
Re-arranging Equation \ref{eq:dft-result}, one finds that the desired information, the complex spectrum $\widetilde{g}(\omega)$ of a Gaussian, can be obtained from $\widetilde{g}(\omega)\simeq e^{-i\omega T_0} \widetilde{g}_k$.

While this specific example discussed the FT of a Gaussian, the Fourier shift theorem applies generally to an arbitrary input pulse $E(t)$, and hence we generalise to write $\widetilde{E}(\omega_k)=e^{-i\omega_k T_0} \mathcal{F}[E(t-T_0)] \simeq e^{-i\omega_k T_0} \widetilde{E}_k.$
This result allows one to derive the phase-corrected DFT of Equation \ref{eq:DFT5} from the phase-ambiguous version in Equation \ref{eq:DFT3}.

\section{Python code}
\label{app:code}

The following Python routines use the numpy module ``fft'' to calculate the complex Fourier transform of the data ``ydata'', which is a function of the time variable ``time''. The complex spectrum as a function of positive and negative frequency is returned.

%\begin{tcolorbox}
\subsection{Forwards Fourier transform}
\begin{verbatim}
from numpy import fft, pi, exp  # import required functions
i=1j                            # define the imaginary number

def takeFFT(time,ydata,T0=0.0e-12):
    n = len(time)
    dt= abs(time[1]-time[0])

    ####### CALCULATE THE DFT USING NUMPY
    fftdata = n * fft.ifft(ydata,n=n)   
    # Notes: - the ifft routine is used as it uses exp(i omega t).
    #        - numpy's ifft divides by n, so need to scale by n to be
    #          able to use this as the forward transform.

    ####### WORK OUT THE FREQUENCY ELEMENTS
    freq=fft.fftfreq(n=n, d=dt)	
    # Notes: - freq runs from 0 -> f_max then -f_max -> -1/T, 
    #          where f_max = 1/dt for time step "dt".

    ####### SHIFT THE DFT
    freq = fft.fftshift(freq)
    fftdata = fft.fftshift(fftdata)
    # Notes: - this makes the sequence run from -f_max to +f_max 
    #          so that it is symmetric about zero frequency, like the CFT

    omega = 2*pi*freq 
    
    ####### CORRECT THE DFT SO THAT IT AGREES WITH THE CFT
    fftdata = fftdata * dt * exp(-i * omega * T0)
    # Notes: - scaling factor "dt" needed for correct amplitude
    #        - phase factor exp(-i * omega * T0) accounts for pulse
    #          centred at time T0
    
    return omega, fftdata
\end{verbatim}
%\end{tcolorbox}

%\begin{tcolorbox}
\subsection{Inverse Fourier transform}
\begin{verbatim}
def takeIFFT(omega,E_omega,T0=0.0e-12):
    delta_omega=omega[1]-omega[0]

    ###### WORK OUT Ek
    Ek = E_omega * exp(i*omega*T0)      
    # Note: - input to DFT routine needs to use standard DFT with relative phase
    
    ###### SHIFT THE DFT 
    Ek = fft.ifftshift(Ek)
    # Note: this puts the sequence back into the right order

    ###### WORK OUT THE TIME-DOMAIN FROM THE INVERSE TRANSFORM AND SCALE IT APPROPRIATELY
    Em = fft.fft(Ek)*delta_omega/(2*pi)
    # Here delta_omega/2 pi = 1/(n dt)
    
    return Em
\end{verbatim}
%\end{tcolorbox}

\section{Detector response function}
\label{app:EOS}
In the time-domain, the detected electro-optic signal $S(t)$ is related to the THz electric field $E(t)$ by $S(t) = R(t) \circledast E(t)$, where the influence of the detection crystal is captured by the convolution with the impulse response function, $R(t)$.
In the frequency domain, via the convolution theorem the electro-optic signal is $\widetilde{S}(\omega) = \widetilde{R}(\omega) \widetilde{E}(\omega)$, and hence we can determine the THz electric field from $\widetilde{E}(\omega)=\widetilde{S}(\omega)/\widetilde{R}(\omega)$.
We used the standard approach to determine $\widetilde{R}(\omega)$: \cite{Casalbuoni2008}
\begin{equation}
\widetilde{R}(\omega) = r_{41}(\omega) \frac{e^{i \Delta k_P L}-1.0}{i \Delta k_P L} \label{eq:EOSresponse} 
\end{equation}
where $\Delta k_P = \omega \left[ \widetilde{n}(\omega)/c - 1/v_g \right]$, for a crystal with thickness $L$, THz complex refractive index $\widetilde{n}$, and a probe beam with group velocity $v_g$. 
This response function includes the velocity mismatch between the THz and probe pulse as a result of the detection crystal's frequency-dependent refractive index \cite{Gallot1999}, as well as the variation in the electro-optic coefficient $r_{41}(\omega)$ with frequency \cite{Leitenstorfer1999}.
Note that because $\widetilde{R}(\omega)$ is complex, and has a phase that is linear in the angular frequency, by the Fourier shift theorem dividing by $\widetilde{R}(\omega)$ changes the arrival time of the THz pulse in the time-domain.
Here we accounted for this shift so that $S(t)$ and $E(t)$ both had $t_e=0$, allowing a more straightforward comparison of the temporal shapes before and after correcting for the detector's response.

\section{Declarations}

\textit{Ethical Approval}\\
N/A.\\ 

\noindent \textit{Acknowledgements}\\
Parts of this research were carried out at ELBE at the Helmholtz-Zentrum Dresden - Rossendorf e.V., a member of the Helmholtz Association. 
The authors would like to thank the TELBE team, in particular A. Arshad, J.-C.\ Deinert, I.\ Ilyakov, A.\ Ponomaryov and G.\ L.\ Prajapati, for support.\\

\noindent \textit{Funding}\\
The authors would like to acknowledge funding from the EPSRC (UK) (Grant No.\ EP/V047914/1) and the European Union’s Horizon Europe research and innovation programme under grants 101131414 (NEPHEWS) and 101103873 (UltraBat). \\

\noindent \textit{Availability of data and materials}\\
Data are available from the authors on reasonable request. \\

% ****** BIBLIOGRAPHY ******
% ******  Important   ******
% Use the following two lines to compile your bibliography from a .bib file, but comment them before submission and copy the content of the .bbl file below before submission. 
%\bibliography{references}
%\bibliographystyle{unsrt}

\end{document}